\documentclass{jpp}
\usepackage{graphicx}
\usepackage{epstopdf, epsfig}
\usepackage{hyperref}
\usepackage{amsmath,amsfonts,amssymb}
\usepackage{subfigure}
\usepackage{tabularx}
\usepackage{comment}
\usepackage{natbib}
\usepackage{xcolor}

\newcommand{\chg}[1]{{#1}}
\shorttitle{CHIMERAS Project Design Framework}
\shortauthor{S. Dorfman, S.~Bose, E.~Lichko, M.~Abler, J.~Juno, J.~M.~TenBarge, et. al.}

\title{The CHIMERAS Project: Design Framework for the Collisionless HIgh-beta Magnetized Experiment Researching Astrophysical Systems}

\author{S.~Dorfman\aff{1,2},
S.~Bose\aff{3,4},
E.~Lichko\aff{5, 6},  M.~Abler\aff{1,2},
J.~Juno\aff{3}, 
J.~M.~TenBarge\aff{4},
Y.~Zhang\aff{4,7},
S.~Chakraborty Thakur\aff{8},
C.~A.~Cartagena-Sanchez\aff{9},
P.~Tatum\aff{10},
E.~Scime\aff{11}, 
G.~Joshi\aff{2},
S.~Greess\aff{12}  \&  \chg{C.~Kuchta\aff{13}}}

\affiliation{\aff{1} Space Science Institute, Boulder, CO, 80301,
    \aff{2} University of California - Los Angeles, Los Angeles, CA, 90095
    \aff{3} Princeton Plasma Physics Laboratory, Princeton, NJ, 08540,
    \aff{4} Princeton University, Princeton, NJ, 08544,
    \aff{5} University of Chicago, Chicago, IL, 60637,
    \aff{6} Naval Research Laboratory, Washington, DC 20375, USA
    \aff{7} University Corporation for Atmospheric Research, Boulder, CO, 80301,
    \aff{8} Auburn University, Auburn, AL, 36849,
    \aff{9} Beloit College, Beloit, WI, 53511,
    \aff{10} University of Colorado, Boulder, CO 80309
    \aff{11} West Virginia University, Morgantown, WV, 26506
    \aff{12} Queen Mary University of London, London E1 4NS, UK
    \chg{\aff{13} University of Wisconsin - Madison, Madison, WI 53715, USA}}

\begin{document}

\maketitle

\begin{abstract}
From the near-Earth solar wind to the intracluster medium of galaxy clusters, collisionless, high-beta, magnetized plasmas pervade our universe.  Energy and momentum transport from large-scale fields and flows to small scale motions of plasma particles is ubiquitous in these systems, but a full picture of the underlying physical mechanisms remains elusive.  The transfer is often mediated by a turbulent cascade of Alfv{\'e}nic fluctuations as well as a variety of kinetic instabilities; these processes tend to be multi-scale and/or multi-dimensional, which makes them difficult to study using spacecraft missions and numerical simulations alone \citep{dorfman23a, lichko2020, Lichko2023}.  Meanwhile, existing laboratory devices struggle to produce the collisionless, high ion beta ($\beta_i \gtrsim 1$), magnetized plasmas across the range of scales necessary to address these problems.  As envisioned in recent community planning documents \citep{carter20, milchberg20, baalrud20, dorfman23a, heliodecadal24}, it is therefore important to build a next generation laboratory facility to create a $\beta_i \gtrsim 1$, collisionless, magnetized plasma in the laboratory for the first time.  A Working Group has been formed and is actively defining the necessary technical requirements to move the facility towards a construction-ready state. Recent progress includes the development of target parameters and diagnostic requirements as well as the identification of a need for source-target device geometry. As the working group is already leading to new synergies across the community, we anticipate a broad community of users funded by a variety of federal agencies (including NASA, DOE, and NSF) to make copious use of the future facility.
\end{abstract}


\section{Motivation and Summary of Goals}

\subsection{Scientific Motivation for a New Facility}\label{sec:goals}


One of the major difficulties in studying space and astrophysical plasmas is characterizing energy and momentum transport across the broad range of scales of dynamical importance. For example, in galactic dynamics, there are roughly twelve orders of magnitude between the ion-scale fluctuations in the galactic disc and the much larger scales on which the energy transfer processes between the thermal plasma and cosmic rays operate \citep{Brunetti2014}. Similar scale problems exist in our local heliosphere, where large-scale fluctuations driven by solar rotation are six orders of magnitude larger than small-scale fluctuations on electron gyration periods that terminate the turbulent cascade \citep{kiyani15}. Understanding the mesoscale processes that determine the energy partition between thermal plasma species and couple the small- and large-scale dynamics is critical for making progress on open questions in both space and astrophysical systems.

The processes that are most critical to understand, and currently most poorly understood, are those occurring in high-beta, magnetized, weakly collisional plasmas \citep{Kunz2019}. These include, but are not limited to: 
\begin{itemize}
\item \textbf{Instabilities} not only produce electromagnetic fluctuations but also constrain these same fluctuations by scattering particles and modifying the plasma's energy and momentum transport.  These processes are important in a variety of astrophysical environments. In the solar wind, the role of pressure-anisotropy driven instabilities in constraining the fluctuations present is well-established \citep{Bale2009}. Outside of the heliosphere, instabilities play a large role in a number of astrophysical systems, including accretion discs at the heart of galaxies, galaxies themselves, and the intracluster medium (ICM) \chg{\citep{Balbus1991, Quataert2002, Quataert2008, Kunz2014, Kunz2022}.} Instabilities also play a large role in cosmic-ray energization near supernova remnants \citep{Bell2004}, the coupling between the thermal plasma in the galactic disc and cosmic rays \citep{Kulsrud1969}, shock heating throughout the universe \citep{Spitkovsky2008, Caprioli2013, Wilson2014, Wilson2016}, and potentially the generation of radio halos \citep{Brunetti2014}. Relevant instabilities include firehose and mirror instabilities \chg{\citep{Gary:2001, Kasper:2002, Hellinger:2006, Schekochihin:2008a, Rosin:2011}}, heat-flux and gradient-driven instabilities \chg{\citep{Riquelme:2016, Komarov:2016, RobergClark:2016, RobergClark:2018, Komarov:2018}}, and streaming instabilities \chg{\citep{Bell2004, Amato2009}}, among others. 
\item \textbf{Turbulence} is a major channel of energy transfer in many space and astrophysical systems.
\chg{Understanding} the details of how turbulence transfers energy from scale to scale \chg{and the processes by which it ultimately transfers energy to the particles are critical for a number of open questions}.
\chg{The resulting energy and momentum transport affects everything from the state of the turbulent solar wind at Earth \citep{breech09,howes10} to the amount and type of radiation emitted from astrophysical objects such as accretion discs \citep{EHT:2019a,EHT:2019b,EHT:2021}. 
Furthermore, turbulence affects the evolution of large-scale structures everywhere from our heliosphere \citep{Tu:1995, Verscharen:2019, Richardson:2022} to more distant galaxies and their surrounding circumgalactic media \citep{Tumlinson:2017, Ji:2020, Lochhaas:2023}. 
Due to the very large scale separation between the turbulent dissipation and the macroscopic evolution, widely used fluid models of these large-scale systems \citep{Toth2012, Thomas2022, talbot2024} use effective heat conduction and viscosity terms to approximate the energy and momentum transport due to turbulent dissipation.  The underlying turbulent dissipation model chosen can significantly affect the observed behavior of the system (e.g.~\citet{Chael:2018, Chael:2019})}.
The ambient\chg{, turbulent} magnetic fluctuations \chg{in these systems} can also affect the trajectory of cosmic rays, the most energetic particles in the universe, and can impact our understanding of where these fast, charged particles originate \citep{Owen2023}.
\end{itemize}

{\bf To enhance our understanding of energy and momentum transport in the aforementioned space and astrophysical systems, a plasma device capable of achieving collisionless conditions while being magnetized with high plasma beta is necessary.} 
\chg{Basic plasma science experiments have had success in studying astrophysical phenomena where one or two of the conditions (collisionless, magnetized, and high beta) are met \citep{Keiter2000, brown_laboratory_2014, Schaffner2014, Dorfman2016, Schroeder2021, Endrizzi:2021, Peterson2021,Ji2023,bose2024experimental}. However,} existing experiments (e.g.,~\citet{gekelman2016upgraded,forest15}) struggle to achieve all three conditions simultaneously, leaving a key gap in our approach to solving the problems mentioned above.
\chg{Unlike spacecraft, laboratory experiments can take multi-point measurements of a controlled, reproducible plasma \citep{Howes2018, lichko2020, Lichko2023}. Such experiments can therefore complement and extend the utility of spacecraft in exploring the multi-scale and multi-dimensional physics that occurs in space and astrophysical systems.} 
Thus, the proposed device will take advantage of these unique capabilities of laboratory experiments to enable novel plasma science experiments with broad relevance to space and astrophysical plasmas.


\subsection{Goals of the Planning Process}\label{sec:requirements}

To tackle this problem, The CHIMERAS (Collisionless HIgh-beta Magnetized Experiment Researching Astrophysical Systems) Project Working Group has been formed under the auspices of MagNetUS, which is a network comprised of several basic plasma science facilities and their collaborators.  The Working Group consists of 30+ scientists from various parts of the community (space observation, numerical simulations, theory, and laboratory experiments) and is actively defining the necessary technical requirements to design an experimental facility to address the science goals in Section~\ref{sec:goals}.  Our planning process has two major aims:

\begin{enumerate}
    \item Design a device to create a $\beta_i = 8 \pi n T_i / B^2 \gtrsim 1$, collisionless, magnetized plasma in the laboratory for the first time.  This new regime will enable novel plasma science experiments with broad relevance to space and astrophysical plasmas.  We will for the first time be able to study astrophysically relevant collisionless instabilities and magnetized plasma turbulence.
    \item Nurture the budding investigator network working on the device by involving researchers from different parts of the field (space observation, computer simulations, theory, and laboratory experiments) and different career stages (including graduate students and postdocs) in the above discussions. This broad array of expertise will expose participants to subject areas and research techniques with which they may not be familiar.
\end{enumerate}

A summary of the key dimensionless parameter requirements necessary to address our first aim is given in Table~\ref{tab:requirements}.  The high beta condition is \chg{necessary because the space and astrophysical environments this machine aims to emulate are typically $\beta_i \gtrsim 1$, and the onset of several of the targeted instabilities is dependent on $\beta_i$}.  The collisionless condition comes from the two-fluid Alfv{\'e}n wave dispersion relation \citep{mallet23}; to explore kinetic scales, we wish to minimize Alfv{\'e}n wave damping in the regime where the waves are dispersive.  \chg{Since the frequency of Alfv{\'e}n waves in the device is expected to range from a fraction of the ion cyclotron frequency ($f_{ci}$) to slightly less than $f_{ci}$, we will sometimes write this criterion in terms of $\nu_{ei}/\omega_{ci}$. The need to fit magnetohydrodynamic (MHD) Alfv{\'e}n waves in the device gives rise to} the magnetized condition; this leads to different conditions for machine length along the background magnetic field $L_\parallel$ and the machine diameter perpendicular to the field $L_\perp$.  The relevant conditions are in terms of the ion gyroradius $\rho_i$ and the ion skin depth $d_i$.  In the perpendicular direction, we need to resolve $k_\perp \rho_i \sim 1$ in order to capture the relevant anisotropy-driven instability physics. Meanwhile, we need to be able to launch Alfv{\'e}n waves at perpendicular scales larger than both $\rho_i$ and $d_i$ in the turbulence experiments in order to avoid complications due to the presence of dispersive effects and Hall effects respectively \citep{mallet23}.  Since the \chg{smallest} $k_\perp$ that can fit in the device will be $2\pi/L_\perp$, the factor of fifty is the minimum necessary to enable both the instability and the turbulence experiments, as it allows for $k_\perp \rho_i \gtrsim 0.13$ (or $k_\perp d_i \gtrsim 0.13$).  The scale requirement in the parallel direction allows a low frequency Alfv{\'e}n wave ($\omega \ll \chg{\omega_{ci}}$) to fit in the device; \chg{this may be seen by noting that the MHD Alfv{\'e}n wave dispersion relation can be written as $k_\parallel d_i = \omega/\omega_{ci}$.  The device will therefore be large enough to contain counter-propagating, low-frequency Alfv{\'e}n waves for the turbulence experiments and large enough to resolve low-frequency Alfv{\'e}n waves that result from the instabilities.}

\begin{table}
  \begin{center}
    \caption{Machine Requirements: key dimensionless parameter requirements that existing facilities struggle to satisfy which will open up a new physical regime for studies of energy and momentum transport via turbulence and instabilities.}
    \label{tab:requirements}
    \begin{tabular}{c|>{\raggedright}p{3.5cm}|>{\raggedright}p{3cm}|p{4cm}} 
       & \textbf{Parameter} & \textbf{Regime} & \textbf{Reason} \\
      \hline\hline
High beta & Ion plasma beta & $\beta_i \gtrsim 1$ & Comparable magnetic and ion kinetic pressure \\
      \hline
Collisionless & Collision frequency / Alfv{\'e}n wave frequency & $\nu_{ei}/\omega < 1$ & Minimize electron-ion collisions \\
      \hline
Magnetized & Machine size in ion scales & $L_\perp \sim 50 \max(d_i,\rho_i)$ $L_\parallel \sim 100 \max(d_i,\rho_i)$ & Study MHD Alfv{\'e}n waves \\
    \end{tabular}
  \end{center}
\end{table}

\section{Accomplishments of the First Workshops}

The CHIMERAS Project Working Group held the first design workshop April 18th-20th, 2024. During the workshop, the working group established a preliminary set of dimensionless parameters; developed a pre-prototype source/target geometry concept; determined measurement and diagnostic requirements; and emphasized the need for training and professional development opportunities for graduate students, post-doctoral scientists, and early-career researchers.  \chg{These parameters and requirements were further refined at the second workshop on December 15th-16th, 2024.}

\subsection{Preliminary Set of Dimensionless Parameters}\label{sec:params}

\begin{figure}
\includegraphics[width=\textwidth]{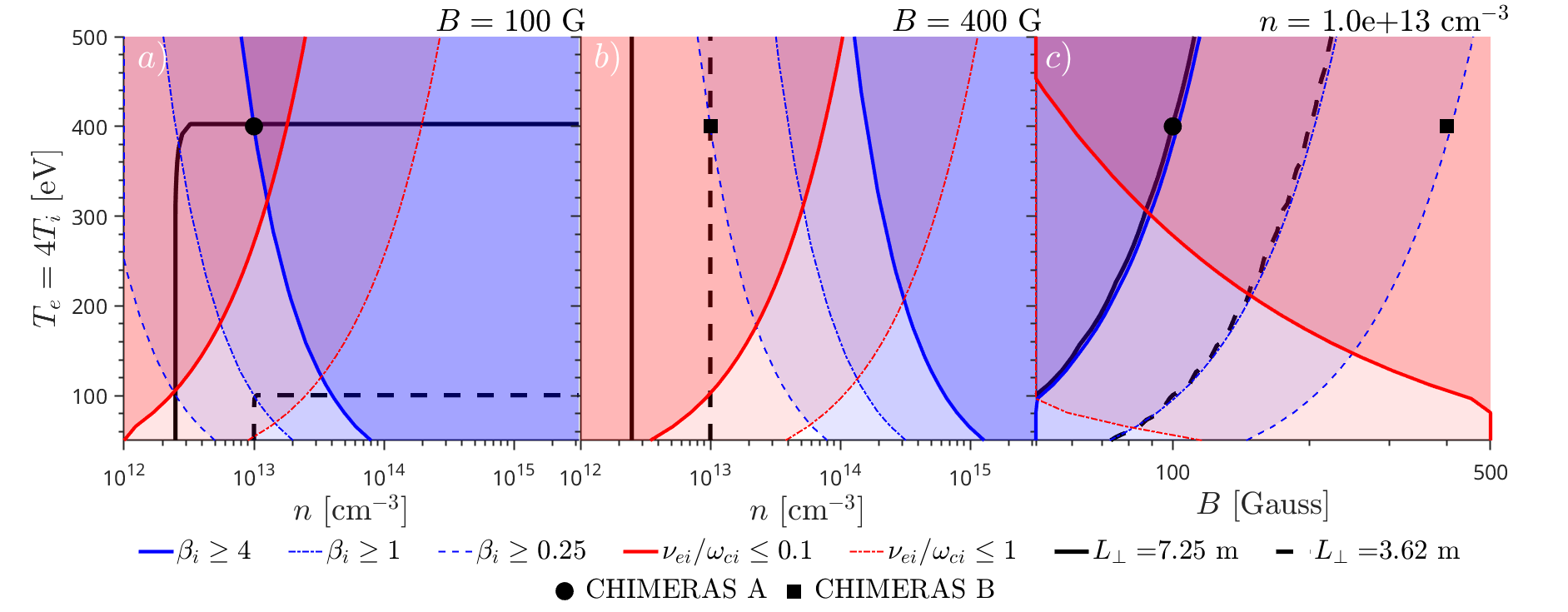}
\caption{\label{fig:paramplot} \chg{Location of Setups A and B in parameter space.  Blue shading indicates different levels of $\beta_i$ while red shading indicates different levels of $\nu_{ei}/\chg{\omega_{ci}}$.  Black lines show the value of $L_\perp=50 \max(d_i,\rho_i)$ for Setup A and B. (a) Location of Setup A in $n$-$T$ parameter space.  (b) Location of Setup B in $n$-$T$ parameter space.  (c) Location of both setups in $B$-$T$ parameter space.  Note that $L_\perp=3.62$m everywhere to the right of the $\beta_i=1$ line in (c), not just on black dashed line.}}
\end{figure}

Our preliminary set of parameters for both the instability (A) and turbulence (B) experiments are outlined in Table~\ref{tab:params1} (dimensional) and Table~\ref{tab:params2} (dimensionless). These parameters are optimized around the requirements in Table~\ref{tab:requirements} and assume that the experiments will be conducted in hydrogen plasma.  \chg{The optimization process is illustrated in Fig.~\ref{fig:paramplot}.  The red and blue shadings indicate different levels of $\beta_i$ and $\nu_{ei}/\omega_{ci}$ respectively; the regions with darker shading are better able to meet each of the first two requirements in Table~\ref{tab:requirements}.} In dimensional terms, the primary difference between the two \chg{experimental setups} is the magnetic field strength, $B_0$; this difference is because the high-beta condition is not strictly necessary for the turbulence experiments (Setup B).

\begin{table}
  \begin{center}
    \caption{Preliminary set of dimensional parameters in hydrogen plasma for both high-beta collisionless instabilities (A) and solar wind magnetized plasma turbulence (B): Parameters are plasma density, electron temperature, ion temperature, background magnetic field, \chg{driven} Alfv\'en wave frequency, \chg{ion cyclotron frequency,} ion skin depth, ion gyroradius, electron-ion collision frequency, \chg{driven} Alfv\'en parallel wavelength, chamber diameter, and chamber length. \chg{Note that the ion cyclotron frequency and Alfv\'en wave frequency are both ordinary frequencies, while the electron-ion collision frequency is an angular frequency.}}
    
    \label{tab:params1}
    \begin{tabular}{l|c|c|c|c|c|c|c|c|c|c|c|c} 
      \textbf{Setup} & \textbf{$n$} & \textbf{$T_e$} & \textbf{$T_i$} &\textbf{$B_0$} &  \textbf{$f_0$}  &  \textbf{$f_{ci}$} &\textbf{$d_i$} & \textbf{$\rho_i$}  & \textbf{$\nu_{ei}$} &\textbf{$\lambda_{\parallel 0}$}\ &\textbf{$L_\perp$} &\textbf{$L_\parallel$} \\ 
       & ($\mathrm{cm}^{-3}$) & (eV) & (eV) & (G)& (kHz) & (kHz)& (cm) & (cm)  & (\chg{rad/s}) & (cm) & (m) & (m) \\
      \hline
      A & $10^{13}$ & $400$ & $100$ & $100$  & $20$ & $152$& $7.21$ & $14.5$  & $54.7$  & $349$ & $7.25$ & $14.5$\\
      B & $10^{13}$ & $400$ & $100$ & $400$  & $75$ & $608$ & $7.21$ & $3.61$  & $54.7$  & $373$ & $3.62$ & $7.25$\\

    \end{tabular}
  \end{center}
\end{table}

\begin{table}
  \begin{center}
    \caption{Preliminary set of dimensionless parameters in hydrogen plasma for both high-beta collisionless instabilities (A) and solar wind magnetized plasma turbulence (B): Parameters are the ratio of electron thermal speed to Alfv{\'e}n speed, parallel wavenumber \chg{corresponding to $f_0$ from Table \ref{tab:params1} }times ion skin depth, the ratio of collisionality to ion cyclotron \chg{angular} frequency $(\chg{\omega_{ci}} =2\pi f_{ci}$), electron beta, and ion beta.}
    \label{tab:params2}
    \begin{tabular}{l|c|c|c|c|c|c} 
      \textbf{Setup} & \textbf{$\frac{{{v_{the}}}}{{{v_A}}}$} & \textbf{$k_{\parallel 0} d_i$} & \textbf{$\frac{{{\nu _{ei}}}}{{{\chg{\omega _{ci}}}}}$} & \textbf{$\beta_e$} & \textbf{$\beta_i$} \\ 
      \hline
      A & $172$ & $0.13$ & $0.057$ & $16.1$ & $4.03$\\
      B & $43.0$ & $0.12$ & $0.014$ & $1.01$ & $0.25$\\

    \end{tabular}
  \end{center}
\end{table}

According to the requirements in Table~\ref{tab:requirements} and parameters in Table~\ref{tab:params1}, a large chamber with diameter $L_ \bot \sim 50\max \left( {{d_i},{\rho _i}} \right)=7.25 ~\rm{m}$ and length $L_ \parallel \sim 100\max \left( {{d_i},{\rho _i}} \right)=14.5 ~\rm{m}$ will work for studying both turbulence as well as kinetic instabilities in a magnetized, collisionless, high ion beta plasma.  Generating high-density plasma ($n > {10^{13}}~{\rm{c}}{{\rm{m}}^{ - 3}}$) in a large chamber is challenging; thus, we use this as our upper limit on plasma density, which effectively sets the lower limit on the device size\footnote{Since $\rho_i \propto \sqrt{\beta_i / n}$ and $d_i \propto 1 / \sqrt{n}$, the required  $\beta_i$ and achievable plasma density $n$ set the values of the ion scales.}. \chg{These limits were calculated assuming a hydrogen plasma}. \chg{A helium plasma would allow for increased diagnostic flexibility (see Section~\ref{subsec:open_diagnostics}), but would increase the scale (in dimensional units) that corresponds to $k_\perp \rho_i = 1$ and thus necessitate a larger vessel size. }

\subsection{Necessity of a Source-Target Geometry}\label{sec:sourcetarget}
To access new plasma parameter regimes in which the plasma is magnetized, collisionless, and high $\beta_i$, it was the conclusion of the workshop discussion that the configuration of the experiment would require a geometry in which a source plasma would fill a target chamber. 
This particular configuration arose from the key constraint in generating both large $\beta_i$ and a magnetized plasma.  Currently operating experiments which have $\beta_i>1$ are too small to achieve the magnetized condition; when the plasma's own magnetic field is weak enough to achieve high beta, the ion gyroradius is often comparable to, or larger than, the system size \citep{Endrizzi:2021, Bott:2021, Meinecke:2022}. Furthermore, any plasma created in a single chamber which has plasma pressure greater than magnetic pressure will rapidly expel the magnetic field; thus, the plasma will not remain magnetized. 

Correspondingly, we must engineer a system in which an initially confined, magnetized, collisionless plasma \emph{source} is allowed to expand into a secondary \emph{target} chamber so that as the plasma expands, the plasma naturally enters the conditions desired due to its own dynamical evolution. \chg{To reduce the loss of plasma to the walls and improve confinement time, the walls of the target chamber can be lined with permanent magnets in either a line-cusp or a broken line-cusp configuration \citep{limpaecher1973magnetic, leung1975optimization, gekelman1975large, forest15}.} 
Additionally, fusion-like temperatures can be realized in the source chamber through combinations of electron cyclotron resonant heating (ECRH) and neutral beam injection (NBI), which would push the plasma \chg{further} into the collisionless regimes necessary to study the kinetic plasmas commonly observed in astrophysical systems. 

This particular configuration has a number of added benefits for the physics goals we seek to address with this new facility. 
For example, the expansion of the plasma in this system is analogous to the expansion of the solar wind. 
This feature will allow the community to study how the effects of expansion \chg{on the distribution function} may be limited by plasma instabilities and how the transport of momentum and heat \chg{in an expanding plasma} may be modified by collisionless processes. 
Further, a source region can be constructed which provides a large degree of flexibility for the temperature of the plasma in the target chamber. 
This flexibility will allow for additional possible diagnostic options. 
For example, the proposed turbulence configuration can tolerate a range of $\beta_i$, so operating at lower temperature (lower $\beta_i$) is feasible. 
\chg{Although this choice will increase the collisionality, Figure~\ref{fig:paramplot} shows there is significant flexibility in maintaining the collisionless condition. 
Decreasing the temperature by a factor of four will still keep the ratio of $\nu_{ei}/\omega_{ci} < 0.1$, thus minimizing the amount of collisional damping expected.}
\subsection{Diagnostic/Measurement Requirements}\label{sec:diagnostics}


 
 The key quantities to be measured are density ($n$), parallel ($\parallel$) and perpendicular ($\perp$) components of electron ($T_{e}$) and ion temperature ($T_{i}$), as well as fluctuations in density ($\delta n$), magnetic field ($\delta B$), and bulk ion velocity ($\delta v$) associated with waves and instabilities.\footnote[2]{\chg{We have also indicated in Table~\ref{tab:diagnostics} when the distribution function ($f$) can be measured. In the indicated diagnostics, the parallel and perpendicular ion and electron temperatures, as well as the velocity fluctuations, are derived from this quantity.}} The fluctuating quantities must be measured with sufficient spatial and temporal resolution to effectively study turbulence and instabilities. For the turbulence experiment, the parallel wavelength ($\lambda_{\parallel}$) of the excited Alfv\'en waves is expected to be comparable to or longer than the perpendicular wavelength ($\lambda_{\perp}$); thus the latter sets the limit for spatial resolution.  For both the turbulence and ion-scale instability experiments, the key kinetic physics, e.g., the largest \chg{ion} damping rates of turbulent fluctuations or the largest growth rates of unstable \chg{ion} modes, is expected to occur around $k_{\perp} \rho_{i} \sim 1$. Therefore, a spatial resolution  smaller than $\sim \rho_{i}$ is desired for Setups A and B. With this resolution and the desired vessel volume (Table~\ref{tab:requirements}), we also have multiple decades of resolved scales for the turbulence studies (from $\sim3.6$ m driving scales down to $\lesssim3.6$ cm sub-$\rho_i$ scales). The time resolution for the  turbulence experiment is set by the requirement to measure oscillating quantities associated with an Alfv\'en wave ($\delta B$, $\delta v$, and at small scales $\delta n$). The frequency of the excited Alfv\'en waves is expected to range from a \chg{fraction of} the ion cyclotron frequency ($f_{ci}$) to slightly less than $f_{ci}$.  For the experiments on anisotropy driven ion instabilities, the bulk of the measurements are expected to be made at $f_{ci}$ scales. Since $f_{ci}$ is the highest frequency that needs to resolved, we require diagnostics that can sample data at rates at least $\sim 10~f_{ci}$. 
 
 Typically, the aforementioned quantities are measured with good spatial and temporal resolution in basic plasma physics experiments using in situ probes. However, we \chg{will} not be able to \chg{extensively} use in-situ probes, as the high heat \chg{and particle} flux \chg{at the densities and} temperatures in our proposed experiments (see Tables~\ref{tab:params1} and \ref{tab:params2}) will significantly reduce the lifetime of the probes. An alternative solution is to adopt optical diagnostics \chg{developed by the fusion community as much as possible. These diagnostics} do not require in-situ components. 



\begin{table}
  \begin{center}
    \caption{Promising  diagnostics that can be used to measure key physical parameters.  Important open questions in the ``Remarks'' column are elaborated on in Section~\ref{subsec:open_diagnostics}.}
    \label{tab:diagnostics}
    \begin{tabular}{c|p{4.5cm}|p{6.5cm}} 
        \textbf{Parameter} & \textbf{Diagnostics} & \textbf{Remarks} \\
\hline \hline
$n$ & Thomson Scattering (TS) \citet{ghazaryan2022thomson,kaur2024design} & Best sampling frequency to be determined\\
 & Multi-Chord Interferometer (MCI) \citet{juhn2021multi,van2004phase} & Line integrated measurement with high sampling frequency\\
\hline
$\chg{f}, T_{e,\parallel}$, $T_{e,\perp}$ &  Thomson Scattering (TS)& Best sampling frequency to be determined \\
 & \citet{shi2023multi} & \\
\hline      
$\delta n/n$ & Beam Emission Spectroscopy (BES)  & Neutral beam is needed. In Tokamaks, BES has been used to measure  2D images of $\delta n/n$ showing turbulent structures. \\
 & \citet{mckee1999beam, kriete2018extracting, bose2022two}  & Spatial resolution may be limited to resolving ion scales. \\ 
 \hline      
$k_{\perp}$ & 2D multichannel Charge Exchange Imaging (CXI) & Neutral beam is needed. Carbon impurities may suffice. \\
 &   &   CXI has a better spatial resolution than BES.   \\
 &  \citet{major2022pedestal}  & We will follow the developments of the fusion community. \\
 \hline
 $k_{\perp}$ &  Fast Camera Imaging (FCI)  &   Previously used to measure the mode structure in a cylindrical plasma.    \\
  &  \citet{thakur2014multi, thakur2014simultaneous}   &    Spatial resolution expected to be better than CXI.   Need to determine the minimum $\delta n/n$ that can be measured.  \\
\hline
$\chg{f}, T_{i,\parallel} , T_{i, \perp}$ & Charge Exchange Recombination Spectroscopy (CHERS) \citet{magee2011anisotropic} & Carbon impurities may suffice\\
 & Laser Induced Fluorescence (LIF) \citet{boivin2003laser, gorbunov2017laser} & May require helium plasmas with helium neutral beam \\
\hline
$\chg{f}, \delta v$ & Laser Induced Fluorescence (LIF) \citet{boivin2003laser} & May require helium plasmas with helium neutral beam\\
\hline
$\delta B$ & Zeeman Quantum Beat Spectroscopy (ZQBS) \citet{test1, gilbert2025non} & Relatively new diagnostic for plasmas. Hydrogen plasmas may not work, and helium or impurities may be an alternative.\\
\hline

    \end{tabular}
  \end{center}


\end{table}

 In fact, because several of the scientific questions CHIMERAS seeks to address require measurements of the particle distribution functions of electrons and ions, optical diagnostics will be a key component of the experimental measurement suite. 
 Techniques such as laser-induced fluorescence \citep {boivin2003laser,gorbunov2017laser} and Thompson scattering \citep{ghazaryan2022thomson,kaur2024design} have been routinely deployed in fusion and other laboratory plasmas, and recent demonstrations of these techniques to reconstruct three-dimensional distribution functions \citep{shi2023multi,gilbert2024lif} are extremely promising. 
 It has now been demonstrated that not only can these diagnostics provide spatially resolved measurements of the bulk non-ideal physics, such as the temperature anisotropy of electrons and ions\footnote[5]{We note that because these measurements are spatially resolved and not line integrated, it need not be the case that the temperature anisotropy be defined with respect to a ``guide'' field and could instead be with respect to a spatially resolved magnetic field perturbation.}, but these diagnostics also offer resolved measurements of detailed non-Maxwellian features of the distribution function \citep{shi2022evdf}. 
 \chg{Measurements of this fidelity allow for detailed phase-space analysis that may be used to identify the details of the energy transfer in collisionless plasmas.  For example, measurements of ion temperature anisotropy in space plasmas have previously been related to ion cyclotron damping \citep{kasper13} and stochastic heating \citep{chandran13}.  Characterization of energy transfer using the field-particle correlation technique \citep{Klein:2016,Howes:2017,Schroeder2021} is a lower priority for CHIMERAS due to the challenges measuring fluctuating electric fields discussed in Section~\ref{subsec:open_diagnostics}}.  

 Table~\ref{tab:diagnostics} shows a list of possible diagnostics that may meet the goals of our experiments after suitable modifications. For some of the diagnostics, such modifications will require further research and development. This point will be discussed in Section~\ref{subsec:open_diagnostics}.

\subsection{Training and Professional Development}

The Working Group also understands the importance of training the next generation of plasma researchers on plasma source development, cutting-edge diagnostics, data analysis techniques, and high-fidelity modeling that are crucial to the broader field of laboratory experiments for space and astrophysical plasmas. This unique facility will bring in scientists from different plasma communities, which will expose students and postdoctoral researchers to a broad range of interdisciplinary plasma topics. This synergy provides the perfect platform to serve as a training ground for young scientists. Moreover, this facility will also serve as a center for extensive plasma outreach and public engagement. The facility will bring together plasma scientists from all parts of the US, including several EPSCoR states, allowing us to implement outreach programs in diverse geographical locations and organize more targeted programs for Primarily Undergraduate Institutes (PUI) and Minority Serving Institutes (MSI), in addition to pursuing Research Experiences for Undergraduates (REU) and Research Experiences for Teachers (RET) programs. Finally, leveraging current working group members' connections, we plan to work with other organizations such as the Coalition for Plasma Science (CPS) and MagNetUS (a network of users of the magnetized plasma Collaborative Research facilities), and the education and outreach committees of the American Physical Society - Division of Plasma Physics (APS - DPP), the Princeton Plasma Physics Laboratory (PPPL), Oak Ridge National laboratory (ORNL), along with federal agencies such as the National Science Foundation (NSF), Department of Energy (DOE), and the National Aeronautics and Space Administration (NASA) to organize summer schools and hands-on training workshops for graduate, undergraduate and high school students and teachers. This unique combination of a cutting-edge research and educational platform will simultaneously serve the missions of workforce development and public engagement.

\section{Outstanding Challenges and Next Steps}

\subsection{Outstanding Challenges}

Discussions at the workshops allowed the working group to more precisely define the outstanding challenges which must be answered before the construction of a device can begin. These questions fall into three categories: \chg{(1) Plasma Evolution and Wave Drive; (2) Diagnostic Development and Choice of Gas Species; and (3) Vessel Size and Shape.}

\subsubsection{Plasma \chg{Evolution} and \chg{Wave} Drive}

\chg{The source-target geometry provides a natural means of attaining conditions similar to the solar wind at 1 AU: $\beta_i \sim 1$, collisionless, and magnetized, but the precise hardware configuration needed to achieve the target parameters listed in Tables 2 and 3 remains unresolved. While the magnetic field can be set by external coils, how the density and temperature evolves following ionization in the source chamber and plasma expansion into the target chamber is an open question. 
It is possible to extend the range of parameters by modifying the plasma once it is within the target chamber. 
This modification could take a number of forms, including neutral beam injection and radio-frequency heating.
The expansion of the plasma can also be further tuned, and thus the excitement of anisotropy-driven instabilities further controlled, in the ``target” part of the source-target geometry by the addition of magnetic coils of variable strength near the interface between the source and target. 
Although the expansion itself may provide a source of turbulence due to the excited instabilities, controlled driving of the turbulence in the desired parameter regime (Setup B in Tables~\ref{tab:params1}~and~\ref{tab:params2}) is desirable. While we may draw inspiration from existing antenna designs (\cite{zhang2008spectral,gigliotti2009generation,thuecks2009tests}), substantial modifications will be necessary to account for the high heat and particle fluxes in CHIMERAS.}  
\chg{Further research is necessary to better understand what hardware and  driving configurations are best suited to our planned studies.}

\subsubsection{Diagnostic Development and Choice of Gas Species}\label{subsec:open_diagnostics}

To measure all the necessary quantities identified in Section \ref{sec:diagnostics} at the required spatial and temporal resolution, further research and diagnostic development will be required. Table~\ref{tab:diagnostics} of Section \ref{sec:diagnostics} gives a list of promising diagnostics that may meet the goals of our experiments after suitable modifications. Some of the questions that need to answered for implementation of the diagnostics are also given in Table~\ref{tab:diagnostics}. In this section, we discuss open challenges that will need to be addressed to successfully measure the fluctuating quantities ($\delta B$, $\delta n$, $\delta v$) identified in Section~\ref{sec:diagnostics} as critical to the facility science goals.  These challenges will require a significant amount of further research.  The solutions will have important implications for the overall device design (see Section~\ref{subsec:sizeshape}) as some measurements may require helium plasma in lieu of the hydrogen setup considered in Tables~\ref{tab:params1} and \ref{tab:params2}.

Among the list of diagnostics in Table~\ref{tab:diagnostics}, Zeeman Quantum Beat Spectroscopy (ZQBS), a method for  measuring magnetic field fluctuations, may require the most significant amount of research  for successful implementation in CHIMERAS \citep{test1, gilbert2025non}. ZQBS will require ion or molecular species that can support Zeeman splitting in the presence of a magnetic field. In a pure hydrogen plasma, the ion species is a proton which will not exhibit the Zeeman effect. However, a diagnostic neutral beam may supply the necessary neutrals that may exhibit the Zeeman effect. Further research is needed to understand if quantum beats can be generated for Zeeman split lines of hydrogen atoms by a commercially available laser. If hydrogen is found to be  unsuitable, then the next option would be to determine if impurities or helium plasma can suffice.      

The outcome of the research on ZQBS will determine how Beam Emission Spectroscopy (BES) is to be implemented in CHIMERAS. BES measures $\delta n/n$ in a 2D plane where $\delta n/n$ can be as low as 0.1\% \citep{mckee1999beam}. BES in fusion devices like DIII-D is used in deuterium plasmas with a deuterium neutral beam \citep{mckee1999beam, bose2022two}. If research on ZQBS suggests that helium ion species are necessary, then we will need to determine suitable helium lines and perform associated atomic physics calculation to support BES in a helium plasma.

To measure ion velocity, Laser Induced Fluorescence (LIF) has  previously been used successfully in helium plasmas \citep{boivin2003laser}, and LIF may also work for measuring velocity fluctuations, e.g.,~\citep{palmer05}. However, if experiments are to be carried out in hydrogen plasmas, then we will need to explore the use of helium impurities to support LIF measurements. 

\chg{Measurement of the wave electric field $\delta E$ using spectroscopic techniques is expected to be difficult in CHIMERAS. Existing techniques used by the fusion community such as the Motional Stark Effect (MSE) \citep{rice1997effect,levinton1999motional} and heavy-ion beam probes \citep{shah1999heavy} have not reported results with the required spatial and temporal resolution outlined in Section~\ref{sec:diagnostics}. We therefore plan to initially rely on the measurement of the quantities in Table~\ref{tab:diagnostics}, with the development of a diagnostic for the measurement of fluctuating electric fields left to future work aimed at expanding the capabilities of the device. } 





\subsubsection{Vessel Size and Shape}\label{subsec:sizeshape}

Any device must be sufficiently large to allow study of the full spatio-temporal evolution of the phenomena in question (see Table \ref{tab:requirements}), which means achieving a quasi-steady state. The minimum vessel size given in  Section~\ref{sec:params} is based on rough estimates of the required spatial scales; however, a more precise quantitative analysis is necessary to move forward with plans for the device. Simulation studies \chg{will help us determine what specific device configurations -  including size, geometry, and species - are needed to achieve and diagnose our target physics.  
Such studies will allow a more precise determination of the expected cost of the experiment, which is strongly dependent on the size of the chamber.}
Per Section~\ref{subsec:open_diagnostics}, running experiments in helium makes additional diagnostic options possible.  We will therefore \chg{also} need to balance the utility of these diagnostic methods against device size considerations in order to determine the best experimental setup for determining the dynamics of high $\beta_i$ instabilities and magnetized plasma turbulence in previously unrealized experimental plasma regimes. 


\subsection{Dissemination of Workshop Results and Next Steps}

The Working Group recognizes the need for community and government support if a device like the one we envision is to become a reality. Our first action toward building the necessary support is to disseminate workshop results to the broader astrophysics, solar physics, heliophysics, and plasma physics communities. Members of the Working Group provided a progress report at the 2024 American Physical Society Division of Plasma Physics conference during the associated MagNetUS reception. As the project matures the Working Group will increase their presence at other conferences, including but not limited to meetings of the American Astronomical Society (AAS), American Geophysical Union (AGU), and Solar Heliospheric and INterplanetary Environment (SHINE) workshop.

The Working Group plans to host semi-annual workshops after the annual MagNetUS and AGU Fall meetings, respectively, to further develop the device concept and increase community engagement with and investment in the project. 
Our next steps are to \chg{further investigate} the current state of the art in plasma diagnostic capabilities and plasma turbulent driver technology \chg{and to design} a pre-prototype device capable of generating and maintaining a collisionless plasma. Potential research projects that support the design of a next-generation device and could be immediately proposed for funding will be discussed; such projects include but are not limited to \chg{diagnostic development efforts (such as those identified in Section~\ref{subsec:open_diagnostics})}, development and testing of drivers for turbulence experiments, and simulations to better establish the physics of the source and plasma expansion into the target chamber to test and refine the device design.
We plan to seek federal and private funding to enable dedicated design and development work and accelerate the current pace of progress.

\section*{Acknowledgements}

The CHIMERAS Project Working Group is currently an underfunded effort; thus we especially appreciate everyone who has contributed to this white paper and/or the April 2024 workshop on a volunteer basis. \chg{This work was supported in part by a workshop grant from the Heising-Simons Foundation \#2024-5662.}  E.~L. was supported in part by NRL base funding and the 2024-2025 Karle Fellowship. J.~M.~T. and S.~B. were supported by NSF award AGS-2401110. Y.~Z. was supported by the NASA Living With a Star Jack Eddy Postdoctoral Fellowship Program, administered by the Cooperative Programs for the Advancement of Earth System Science (CPAESS) under award $\#$80NSSC22M0097.

\section*{Declaration of Interests}
The authors report no conflict of interest.

\renewcommand{\refname}{References}

\bibliographystyle{plainnat}


\begin{thebibliography}{97}
\providecommand{\natexlab}[1]{#1}
\providecommand{\url}[1]{\texttt{#1}}
\expandafter\ifx\csname urlstyle\endcsname\relax
  \providecommand{\doi}[1]{doi: #1}\else
  \providecommand{\doi}{doi: \begingroup \urlstyle{rm}\Url}\fi

\bibitem[Amato and Blasi(2009)]{Amato2009}
E.~Amato and P.~Blasi.
\newblock A kinetic approach to cosmic-ray-induced streaming instability at supernova shocks.
\newblock \emph{Monthly Notices of the Royal Astronomical Society}, 392:\penalty0 1591--1600, 2009.
\newblock ISSN 13652966.
\newblock \doi{10.1111/j.1365-2966.2008.14200.x}.

\bibitem[Baalrud et~al.(2020)Baalrud, Ferraro, Garrison, Howard, Kuranz, Sarff, and Solomon]{baalrud20}
Scott Baalrud, Nathaniel Ferraro, Lauren Garrison, Nathan Howard, Carolyn Kuranz, John Sarff, and Wayne Solomon.
\newblock A community plan for fusion energy and discovery plasma sciences.
\newblock \emph{arXiv preprint arXiv:2011.04806}, 2020.
\newblock https://arxiv.org/abs/2011.04806.

\bibitem[Balbus and Hawley(1991)]{Balbus1991}
Steven~A Balbus and John~F Hawley.
\newblock {A Powerful Local Shear Instability in Weakly Magnetized Disks. I. Linear Analysis}.
\newblock \emph{ApJ}, 376\penalty0 (Ii):\penalty0 214--222, 1991.
\newblock URL \url{http://articles.adsabs.harvard.edu/cgi-bin/nph-iarticle_query?1991ApJ...376..214B&data_type=PDF_HIGH&whole_paper=YES&type=PRINTER&filetype=.pdf}.

\bibitem[Bale et~al.(2009)Bale, Kasper, Howes, Quataert, Salem, and Sundkvist]{Bale2009}
S.~D. Bale, J.~C. Kasper, G.~G. Howes, E.~Quataert, C.~Salem, and D.~Sundkvist.
\newblock {Magnetic Fluctuation Power Near Proton Temperature Anisotropy Instability Thresholds in the Solar Wind}.
\newblock \emph{Physical Review Letters}, 2009.
\newblock ISSN 00319007.
\newblock \doi{10.1103/PhysRevLett.103.211101}.

\bibitem[Bell(2004)]{Bell2004}
A.~R. Bell.
\newblock {Turbulent amplification of magnetic field and diffusive shock acceleration of cosmic rays}.
\newblock \emph{Monthly Notices of the Royal Astronomical Society}, 353\penalty0 (2):\penalty0 550--558, 2004.
\newblock ISSN 00358711.
\newblock \doi{10.1111/j.1365-2966.2004.08097.x}.

\bibitem[Boivin and Scime(2003)]{boivin2003laser}
RF~Boivin and EE~Scime.
\newblock Laser induced fluorescence in ar and he plasmas with a tunable diode laser.
\newblock \emph{Review of scientific instruments}, 74\penalty0 (10):\penalty0 4352--4360, 2003.

\bibitem[Bose et~al.(2022)Bose, Fox, Liu, Yan, McKee, Goodman, and Ji]{bose2022two}
Sayak Bose, William Fox, Dingyun Liu, Zheng Yan, George McKee, Aaron Goodman, and Hantao Ji.
\newblock Two-dimensional plasma density evolution local to the inversion layer during sawtooth crash events using beam emission spectroscopy.
\newblock \emph{Review of Scientific Instruments}, 93\penalty0 (9), 2022.

\bibitem[Bose et~al.(2024)Bose, TenBarge, Carter, Hahn, Ji, Juno, Savin, Tripathi, and Vincena]{bose2024experimental}
Sayak Bose, Jason~M TenBarge, Troy Carter, Michael Hahn, Hantao Ji, James Juno, Daniel~Wolf Savin, Shreekrishna Tripathi, and Stephen Vincena.
\newblock Experimental study of alfv{\'e}n wave reflection from an alfv{\'e}n-speed gradient relevant to the solar coronal holes.
\newblock \emph{The Astrophysical Journal}, 971\penalty0 (1):\penalty0 72, 2024.

\bibitem[Bott et~al.(2021)Bott, Tzeferacos, Chen, Palmer, Rigby, Bell, Bingham, Birkel, Graziani, Froula, Katz, Koenig, Kunz, Li, Meinecke, Miniati, Petrasso, Park, Remington, Reville, Ross, Ryu, Ryutov, Séguin, White, Schekochihin, Lamb, and Gregori]{Bott:2021}
Archie F.~A. Bott, Petros Tzeferacos, Laura Chen, Charlotte A.~J. Palmer, Alexandra Rigby, Anthony~R. Bell, Robert Bingham, Andrew Birkel, Carlo Graziani, Dustin~H. Froula, Joseph Katz, Michel Koenig, Matthew~W. Kunz, Chikang Li, Jena Meinecke, Francesco Miniati, Richard Petrasso, Hye-Sook Park, Bruce~A. Remington, Brian Reville, J.~Steven Ross, Dongsu Ryu, Dmitri Ryutov, Fredrick~H. Séguin, Thomas~G. White, Alexander~A. Schekochihin, Donald~Q. Lamb, and Gianluca Gregori.
\newblock Time-resolved turbulent dynamo in a laser plasma.
\newblock \emph{Proceedings of the National Academy of Sciences}, 118\penalty0 (11):\penalty0 e2015729118, 2021.
\newblock \doi{10.1073/pnas.2015729118}.
\newblock URL \url{https://www.pnas.org/doi/abs/10.1073/pnas.2015729118}.

\bibitem[Breech et~al.(2009)Breech, Matthaeus, Cranmer, Kasper, and Oughton]{breech09}
B.~Breech, W.~H. Matthaeus, S.~R. Cranmer, J.~C. Kasper, and S.~Oughton.
\newblock Electron and proton heating by solar wind turbulence.
\newblock \emph{Journal of Geophysical Research: Space Physics}, 114\penalty0 (A9), 2009.
\newblock \doi{https://doi.org/10.1029/2009JA014354}.
\newblock URL \url{https://agupubs.onlinelibrary.wiley.com/doi/abs/10.1029/2009JA014354}.

\bibitem[Brown and Schaffner(2014)]{brown_laboratory_2014}
M~R Brown and D~A Schaffner.
\newblock Laboratory sources of turbulent plasma: a unique {MHD} plasma wind tunnel.
\newblock \emph{Plasma Sources Science and Technology}, 23\penalty0 (6):\penalty0 063001, September 2014.
\newblock ISSN 0963-0252, 1361-6595.
\newblock \doi{10.1088/0963-0252/23/6/063001}.
\newblock URL \url{https://iopscience.iop.org/article/10.1088/0963-0252/23/6/063001}.

\bibitem[Brunetti and Jones(2014)]{Brunetti2014}
Gianfranco Brunetti and Thomas~W. Jones.
\newblock {Cosmic rays in galaxy clusters and their nonthermal emission}.
\newblock \emph{International Journal of Modern Physics D}, 23\penalty0 (4):\penalty0 1--76, 2014.
\newblock ISSN 02182718.
\newblock \doi{10.1142/S0218271814300079}.

\bibitem[Caprioli and Spitkovsky(2013)]{Caprioli2013}
D.~Caprioli and A.~Spitkovsky.
\newblock {Cosmic-ray-induced filamentation instability in collisionless shocks}.
\newblock \emph{Astrophysical Journal Letters}, 765\penalty0 (1), 2013.
\newblock ISSN 20418205.
\newblock \doi{10.1088/2041-8205/765/1/L20}.

\bibitem[Carter et~al.(2020)Carter, Baalrud, Betti, Ellis, Foster, Geddes, Gleason, Holland, Humrickhouse, Kessel, et~al.]{carter20}
Troy Carter, Scott Baalrud, Riccardo Betti, Tyler Ellis, John Foster, Cameron Geddes, Arianna Gleason, Chistopher Holland, Paul Humrickhouse, Charles Kessel, et~al.
\newblock Powering the future: Fusion \& plasmas.
\newblock Technical report, US Department of Energy (USDOE), Washington, DC (United States). Office of Science and Technical Information, 2020.

\bibitem[Chael et~al.(2018)Chael, Rowan, Narayan, Johnson, and Sironi]{Chael:2018}
Andrew Chael, Michael Rowan, Ramesh Narayan, Michael Johnson, and Lorenzo Sironi.
\newblock {The role of electron heating physics in images and variability of the Galactic Centre black hole Sagittarius A*}.
\newblock \emph{Monthly Notices of the Royal Astronomical Society}, 478\penalty0 (4):\penalty0 5209--5229, 06 2018.
\newblock ISSN 0035-8711.
\newblock \doi{10.1093/mnras/sty1261}.
\newblock URL \url{https://doi.org/10.1093/mnras/sty1261}.

\bibitem[Chael et~al.(2019)Chael, Narayan, and Johnson]{Chael:2019}
Andrew Chael, Ramesh Narayan, and Michael~D Johnson.
\newblock {Two-temperature, Magnetically Arrested Disc simulations of the jet from the supermassive black hole in M87}.
\newblock \emph{Monthly Notices of the Royal Astronomical Society}, 486\penalty0 (2):\penalty0 2873--2895, 04 2019.
\newblock ISSN 0035-8711.
\newblock \doi{10.1093/mnras/stz988}.
\newblock URL \url{https://doi.org/10.1093/mnras/stz988}.

\bibitem[Chandran et~al.(2013)Chandran, Verscharen, Quataert, Kasper, Isenberg, and Bourouaine]{chandran13}
B.~D.~G. Chandran, D.~Verscharen, E.~Quataert, J.~C. Kasper, P.~A. Isenberg, and S.~Bourouaine.
\newblock Stochastic heating, differential flow, and the alpha-to-proton temperature ratio in the solar wind.
\newblock \emph{The Astrophysical Journal}, 776\penalty0 (1):\penalty0 45, sep 2013.
\newblock \doi{10.1088/0004-637X/776/1/45}.
\newblock URL \url{https://dx.doi.org/10.1088/0004-637X/776/1/45}.

\bibitem[Dorfman and Carter(2016)]{Dorfman2016}
S.~Dorfman and T.~A. Carter.
\newblock {Observation of an Alfv{\'{e}}n Wave Parametric Instability in a Laboratory Plasma}.
\newblock \emph{Physical Review Letters}, 116\penalty0 (19):\penalty0 1--5, 2016.
\newblock ISSN 10797114.
\newblock \doi{10.1103/PhysRevLett.116.195002}.

\bibitem[Dorfman et~al.(2023)Dorfman, Lichko, Olson, Juno, Kostadinova, Schaffner, Abler, Thakur, Heuer, Mallet, Li, Howes, Squire, Endrizzi, Young, Schaeffer, Klein, Filwett, Rivera, Guidoni, Timm, TenBarge, Matthews, Arzamasskiy, Du, Comisso, Effenberg, Fries, Shi, Verniero, Ofman, Meyrand, Moreland, Wang, Adhikari, Ledvina, Cranmer, Dong, Gilly, Ghadjari, Juie, Light, Sarkar, Liu, Swisdak, Lynch, Maharana, Fu, Wanliss, Kumar, Kumari, and Preisser]{dorfman23a}
Seth Dorfman, Emily Lichko, Joseph Olson, James Juno, Evdokiya Kostadinova, David Schaffner, Mel Abler, Saikat~Chakraborty Thakur, Peter Heuer, Alfred Mallet, Feiyu Li, Gregory~G. Howes, Jonathan Squire, Douglass Endrizzi, Rachel Young, Derek Schaeffer, Kristopher Klein, Rachael Filwett, Yeimy Rivera, Silvina Guidoni, Arian Timm, Jason TenBarge, Lorin Matthews, Lev Arzamasskiy, Tiger Du, Luca Comisso, Florian Effenberg, Dan Fries, Peiyun Shi, Jaye Verniero, Leon Ofman, Romain Meyrand, Kimberly Moreland, Liang Wang, Subash Adhikari, Vincent Ledvina, Steven Cranmer, Chuanfei Dong, Chris Gilly, Hossein Ghadjari, Shetye Juie, Christopher Light, Ranadeep Sarkar, Yi-Hsin Liu, Marc Swisdak, Benjamin~J. Lynch, Anwesha Maharana, Xiangrong Fu, James Wanliss, Pankaj Kumar, Anshu Kumari, and Luis Preisser.
\newblock Next {{Generation Machine}} to {{Study Heliophysics}} in the {{Laboratory}}.
\newblock \emph{Bulletin of the AAS}, 55\penalty0 (3), July 2023.

\bibitem[Endrizzi et~al.(2021)Endrizzi, Egedal, Clark, Flanagan, Greess, Milhone, Millet-Ayala, Olson, Peterson, Wallace, and Forest]{Endrizzi:2021}
Douglass Endrizzi, J.~Egedal, M.~Clark, K.~Flanagan, S.~Greess, J.~Milhone, A.~Millet-Ayala, J.~Olson, E.~E. Peterson, J.~Wallace, and C.~B. Forest.
\newblock Laboratory resolved structure of supercritical perpendicular shocks.
\newblock \emph{Phys. Rev. Lett.}, 126:\penalty0 145001, Apr 2021.
\newblock \doi{10.1103/PhysRevLett.126.145001}.
\newblock URL \url{https://link.aps.org/doi/10.1103/PhysRevLett.126.145001}.

\bibitem[Forest et~al.(2015)Forest, Flanagan, Brookhart, Clark, Cooper, D{\'e}sangles, Egedal, Endrizzi, Khalzov, Li, and {al.}]{forest15}
C.~B. Forest, K.~Flanagan, M.~Brookhart, M.~Clark, C.~M. Cooper, V.~D{\'e}sangles, J.~Egedal, D.~Endrizzi, I.~V. Khalzov, H.~Li, and et~{al.}
\newblock The wisconsin plasma astrophysics laboratory.
\newblock \emph{Journal of Plasma Physics}, 81\penalty0 (5):\penalty0 345810501, 2015.
\newblock \doi{10.1017/S0022377815000975}.

\bibitem[Gary et~al.(2001)Gary, Skoug, Steinberg, and Smith]{Gary:2001}
S.~Peter Gary, Ruth~M. Skoug, John~T. Steinberg, and Charles~W. Smith.
\newblock Proton temperature anisotropy constraint in the solar wind: Ace observations.
\newblock \emph{Geophys.~Res.~Lett.}, 28\penalty0 (14):\penalty0 2759--2762, 2001.
\newblock \doi{https://doi.org/10.1029/2001GL013165}.
\newblock URL \url{https://agupubs.onlinelibrary.wiley.com/doi/abs/10.1029/2001GL013165}.

\bibitem[Gekelman and Stenzel(1975)]{gekelman1975large}
W~Gekelman and RL~Stenzel.
\newblock Large, quiescent, magnetized plasma for wave studies.
\newblock \emph{Review of Scientific Instruments}, 46\penalty0 (10):\penalty0 1386--1393, 1975.

\bibitem[Gekelman et~al.(2016)Gekelman, Pribyl, Lucky, Drandell, Leneman, Maggs, Vincena, Van~Compernolle, Tripathi, Morales, et~al.]{gekelman2016upgraded}
W~Gekelman, P~Pribyl, Z~Lucky, M~Drandell, D~Leneman, J~Maggs, S~Vincena, B~Van~Compernolle, SKP Tripathi, G~Morales, et~al.
\newblock The upgraded large plasma device, a machine for studying frontier basic plasma physics.
\newblock \emph{Review of Scientific Instruments}, 87\penalty0 (2):\penalty0 025105, 2016.
\newblock \doi{10.1063/1.4941079}.

\bibitem[Ghazaryan et~al.(2022)Ghazaryan, Kaloyan, Gekelman, Lucky, Vincena, Tripathi, Pribyl, and Niemann]{ghazaryan2022thomson}
Sofiya Ghazaryan, Marietta Kaloyan, Walter Gekelman, Zoltan Lucky, Stephen Vincena, SKP Tripathi, P~Pribyl, and C~Niemann.
\newblock Thomson scattering on the large plasma device.
\newblock \emph{Review of Scientific Instruments}, 93\penalty0 (8), 2022.

\bibitem[Gigliotti et~al.(2009)Gigliotti, Gekelman, Pribyl, Vincena, Karavaev, Shao, Sharma, and Papadopoulos]{gigliotti2009generation}
Alex Gigliotti, Walter Gekelman, Patrick Pribyl, Stephen Vincena, Alex Karavaev, Xi~Shao, A~Surjalal Sharma, and D~Papadopoulos.
\newblock Generation of polarized shear alfv{\'e}n waves by a rotating magnetic field source.
\newblock \emph{Physics of Plasmas}, 16\penalty0 (9), 2009.

\bibitem[Gilbert et~al.(2024)Gilbert, Steinberger, and Scime]{gilbert2024lif}
T.~J. Gilbert, T.~E. Steinberger, and E.~E. Scime.
\newblock {Improving pulsed laser induced fluorescence distribution function analysis through matched filter signal processing}.
\newblock \emph{Review of Scientific Instruments}, 95\penalty0 (8):\penalty0 083521, 08 2024.
\newblock ISSN 0034-6748.
\newblock \doi{10.1063/5.0215510}.
\newblock URL \url{https://doi.org/10.1063/5.0215510}.

\bibitem[Gilbert et~al.(2025)Gilbert, Steinberger, and Scime]{gilbert2025non}
Tyler~James Gilbert, Thomas Steinberger, and Earl~E Scime.
\newblock Non-intrusive measurement of magnetic field strengths in a low-pressure argon plasma using quantum beat spectroscopy.
\newblock \emph{Plasma Sources Science and Technology}, 2025.

\bibitem[Gorbunov et~al.(2017)Gorbunov, Mukhin, Berik, Vukolov, Lisitsa, Kukushkin, Levashova, Barnsley, Vayakis, and Walsh]{gorbunov2017laser}
AV~Gorbunov, EE~Mukhin, EB~Berik, K~Yu Vukolov, VS~Lisitsa, AS~Kukushkin, MG~Levashova, R~Barnsley, G~Vayakis, and MJ~Walsh.
\newblock Laser-induced fluorescence for iter divertor plasma.
\newblock \emph{Fusion Engineering and Design}, 123:\penalty0 695--698, 2017.

\bibitem[{Hellinger} et~al.(2006){Hellinger}, {Tr{\'a}vn{\'{\i}}{\v c}ek}, {Kasper}, and {Lazarus}]{Hellinger:2006}
P.~{Hellinger}, P.~{Tr{\'a}vn{\'{\i}}{\v c}ek}, J.~C. {Kasper}, and A.~J. {Lazarus}.
\newblock {Solar wind proton temperature anisotropy: Linear theory and WIND/SWE observations}.
\newblock \emph{Geophys.~Res.~Lett.}, 33:\penalty0 9101--+, May 2006.
\newblock \doi{10.1029/2006GL025925}.

\bibitem[Howes(2010)]{howes10}
G.~G. Howes.
\newblock {A prescription for the turbulent heating of astrophysical plasmas}.
\newblock \emph{Monthly Notices of the Royal Astronomical Society: Letters}, 409\penalty0 (1):\penalty0 L104--L108, 11 2010.
\newblock ISSN 1745-3925.
\newblock \doi{10.1111/j.1745-3933.2010.00958.x}.
\newblock URL \url{https://doi.org/10.1111/j.1745-3933.2010.00958.x}.

\bibitem[Howes(2018)]{Howes2018}
Gregory~G. Howes.
\newblock Laboratory space physics: Investigating the physics of space plasmas in the laboratory.
\newblock \emph{Physics of Plasmas}, 25, 5 2018.
\newblock ISSN 10897674.
\newblock \doi{10.1063/1.5025421}.

\bibitem[Howes et~al.(2017)Howes, Klein, and Li]{Howes:2017}
Gregory~G. Howes, Kristopher~G. Klein, and Tak~Chu Li.
\newblock Diagnosing collisionless energy transfer using field–particle correlations: Vlasov–poisson plasmas.
\newblock \emph{J.~Plasma Phys.}, 83\penalty0 (1):\penalty0 705830102, 2017.
\newblock \doi{10.1017/S0022377816001197}.
\newblock URL \url{https://www.cambridge.org/core/article/diagnosing-collisionless-energy-transfer-using-fieldparticle-correlations-vlasovpoisson-plasmas/12921EF6C25C03AC133C892D5A2E4B89}.

\bibitem[Ji et~al.(2023)Ji, Yoo, Fox, Yamada, Argall, Egedal, Liu, Wilder, Eriksson, Daughton, Bergstedt, Bose, Burch, Torbert, Ng, and Chen]{Ji2023}
H.~Ji, J.~Yoo, W.~Fox, M.~Yamada, M.~Argall, J.~Egedal, Y.~H. Liu, R.~Wilder, S.~Eriksson, W.~Daughton, K.~Bergstedt, S.~Bose, J.~Burch, R.~Torbert, J.~Ng, and L.~J. Chen.
\newblock Laboratory study of collisionless magnetic reconnection, 12 2023.
\newblock ISSN 15729672.

\bibitem[Ji et~al.(2020)Ji, Chan, Hummels, Hopkins, Stern, Kereš, Quataert, Faucher-Giguère, and Murray]{Ji:2020}
Suoqing Ji, T~K Chan, Cameron~B Hummels, Philip~F Hopkins, Jonathan Stern, Dušan Kereš, Eliot Quataert, Claude-André Faucher-Giguère, and Norman Murray.
\newblock Properties of the circumgalactic medium in cosmic ray-dominated galaxy haloes.
\newblock \emph{Mon.~Not.~Roy.~Astron.~Soc.}, 496\penalty0 (4):\penalty0 4221--4238, 06 2020.
\newblock ISSN 0035-8711.
\newblock \doi{10.1093/mnras/staa1849}.
\newblock URL \url{https://doi.org/10.1093/mnras/staa1849}.

\bibitem[Juhn et~al.(2021)Juhn, Lee, Lee, Wi, Kim, Hahn, and Nam]{juhn2021multi}
June-Woo Juhn, KC~Lee, TG~Lee, HM~Wi, YS~Kim, SH~Hahn, and YU~Nam.
\newblock Multi-chord ir--visible two-color interferometer on kstar.
\newblock \emph{Review of Scientific Instruments}, 92\penalty0 (4), 2021.

\bibitem[{Kasper} et~al.(2002){Kasper}, {Lazarus}, and {Gary}]{Kasper:2002}
J.~C. {Kasper}, A.~J. {Lazarus}, and S.~P. {Gary}.
\newblock {Wind/SWE observations of firehose constraint on solar wind proton temperature anisotropy}.
\newblock \emph{Geophys.~Res.~Lett.}, 29:\penalty0 20--1, September 2002.

\bibitem[Kasper et~al.(2013)Kasper, Maruca, Stevens, and Zaslavsky]{kasper13}
Justin~C. Kasper, Bennett~A. Maruca, Michael~L. Stevens, and Arnaud Zaslavsky.
\newblock Sensitive test for ion-cyclotron resonant heating in the solar wind.
\newblock \emph{Phys. Rev. Lett.}, 110:\penalty0 091102, Feb 2013.
\newblock \doi{10.1103/PhysRevLett.110.091102}.
\newblock URL \url{https://link.aps.org/doi/10.1103/PhysRevLett.110.091102}.

\bibitem[Kaur et~al.(2024)Kaur, Diallo, LeBlanc, Segado-Fernandez, Viezzer, Huxford, Mancini, Cruz-Zabala, Podesta, Berkery, et~al.]{kaur2024design}
M~Kaur, A~Diallo, B~LeBlanc, J~Segado-Fernandez, E~Viezzer, RB~Huxford, A~Mancini, DJ~Cruz-Zabala, M~Podesta, JW~Berkery, et~al.
\newblock Design of a thomson scattering diagnostic for the small aspect ratio tokamak (smart).
\newblock \emph{Review of Scientific Instruments}, 95\penalty0 (9), 2024.

\bibitem[Keiter et~al.(2000)Keiter, Scime, Balkey, Boivin, Kline, and Gary]{Keiter2000}
Paul~A. Keiter, Earl~E. Scime, Matthew~M. Balkey, Robert Boivin, John~L. Kline, and S.~Peter Gary.
\newblock Beta-dependent upper bound on ion temperature anisotropy in a laboratory plasma.
\newblock \emph{Physics of Plasmas}, 7\penalty0 (3):\penalty0 779--783, 03 2000.
\newblock ISSN 1070-664X.
\newblock \doi{10.1063/1.873872}.
\newblock URL \url{https://doi.org/10.1063/1.873872}.

\bibitem[Kiyani et~al.(2015)Kiyani, Osman, and Chapman]{kiyani15}
Khurom~H. Kiyani, Kareem~T. Osman, and Sandra~C. Chapman.
\newblock Dissipation and heating in solar wind turbulence: from the macro to the micro and back again.
\newblock \emph{Philosophical Transactions of the Royal Society A: Mathematical, Physical and Engineering Sciences}, 373\penalty0 (2041):\penalty0 20140155, 2015.
\newblock \doi{10.1098/rsta.2014.0155}.
\newblock URL \url{https://royalsocietypublishing.org/doi/abs/10.1098/rsta.2014.0155}.

\bibitem[Klein and Howes(2016)]{Klein:2016}
K.~G. Klein and G.~G. Howes.
\newblock Measuring collisionless damping in heliospheric plasmas using field--particle correlations.
\newblock \emph{Astrophys.~J.~Lett.}, 826\penalty0 (2):\penalty0 L30, jul 2016.
\newblock \doi{10.3847/2041-8205/826/2/l30}.
\newblock URL \url{https://doi.org/10.3847%2F2041-8205%2F826%2F2%2Fl30}.

\bibitem[Komarov et~al.(2018)Komarov, Schekochihin, Churazov, and Spitkovsky]{Komarov:2018}
S.~Komarov, A.~A. Schekochihin, E.~Churazov, and A.~Spitkovsky.
\newblock Self-inhibiting thermal conduction in a high- $\beta$ , whistler-unstable plasma.
\newblock \emph{J.~Plasma Phys.}, 84\penalty0 (3):\penalty0 905840305, 2018.
\newblock \doi{10.1017/S0022377818000399}.

\bibitem[Komarov et~al.(2016)Komarov, Churazov, Kunz, and Schekochihin]{Komarov:2016}
S.~V. Komarov, E.~M. Churazov, M.~W. Kunz, and A.~A. Schekochihin.
\newblock {Thermal conduction in a mirror-unstable plasma}.
\newblock \emph{Mon.~Not.~Roy.~Astron.~Soc.}, 460\penalty0 (1):\penalty0 467--477, 04 2016.
\newblock ISSN 0035-8711.
\newblock \doi{10.1093/mnras/stw963}.
\newblock URL \url{https://doi.org/10.1093/mnras/stw963}.

\bibitem[Kriete et~al.(2018)Kriete, McKee, Fonck, Smith, Whelan, and Yan]{kriete2018extracting}
DM~Kriete, GR~McKee, RJ~Fonck, DR~Smith, GG~Whelan, and Z~Yan.
\newblock Extracting the turbulent flow-field from beam emission spectroscopy images using velocimetry.
\newblock \emph{Review of Scientific Instruments}, 89\penalty0 (10), 2018.

\bibitem[Kulsrud and Pearce(1969)]{Kulsrud1969}
Russell Kulsrud and William~P. Pearce.
\newblock {The Effect of Wave-Particle Interactions on the Propagation of Cosmic Rays}.
\newblock \emph{The Astrophysical Journal}, 156\penalty0 (5):\penalty0 445--469, 1969.

\bibitem[Kunz et~al.(2019)Kunz, Squire, Balbus, Bale, Chen, Churazov, Cowley, Forest, Gammie, Quataert, Reynolds, Schekochihin, Sironi, Spitkovsky, Stone, Zhuravleva, and Zweibel]{Kunz2019}
M.~W. Kunz, J.~Squire, S.~A. Balbus, S.~D. Bale, C.~H.~K. Chen, E.~Churazov, S.~C. Cowley, C.~B. Forest, C.~F. Gammie, E.~Quataert, C.~S. Reynolds, A.~A. Schekochihin, L.~Sironi, A.~Spitkovsky, J.~M. Stone, I.~Zhuravleva, and E.~G. Zweibel.
\newblock {[Plasma 2020 Decadal] The Material Properties of Weakly Collisional, High-Beta Plasmas}.
\newblock \emph{arXiv preprint arXiv:1903.04080}, 2019.

\bibitem[Kunz et~al.(2014)Kunz, Schekochihin, and Stone]{Kunz2014}
Matthew~W. Kunz, Alexander~A. Schekochihin, and James~M. Stone.
\newblock {Firehose and mirror instabilities in a collisionless shearing plasma}.
\newblock \emph{Physical Review Letters}, 112\penalty0 (20):\penalty0 1--6, 2014.
\newblock ISSN 10797114.
\newblock \doi{10.1103/PhysRevLett.112.205003}.

\bibitem[Kunz et~al.(2022)Kunz, Jones, and Zhuravleva]{Kunz2022}
Matthew~W. Kunz, Thomas~W. Jones, and Irina Zhuravleva.
\newblock \emph{Plasma Physics of the Intracluster Medium}, pages 1--42.
\newblock Springer Nature Singapore, Singapore, 2022.
\newblock ISBN 978-981-16-4544-0.
\newblock \doi{10.1007/978-981-16-4544-0_125-1}.
\newblock URL \url{https://doi.org/10.1007/978-981-16-4544-0_125-1}.

\bibitem[Leung et~al.(1975)Leung, Samec, and Lamm]{leung1975optimization}
KN~Leung, TK~Samec, and A~Lamm.
\newblock Optimization of permanent magnet plasma confinement.
\newblock \emph{Physics Letters A}, 51\penalty0 (8):\penalty0 490--492, 1975.

\bibitem[Levinton(1999)]{levinton1999motional}
FM~Levinton.
\newblock The motional stark effect: Overview and future development.
\newblock \emph{Review of scientific instruments}, 70\penalty0 (1):\penalty0 810--814, 1999.

\bibitem[Lichko et~al.(2020)Lichko, Endrizzi, Juno, Olson, Dorfman, and Young]{lichko2020}
E.~Lichko, D.~Endrizzi, J.~Juno, J.~Olson, S.~Dorfman, and R.~Young.
\newblock {Enabling Discoveries in Heliospheric Science through Laboratory Plasma Experiments [White paper for the Heliophysics 2050 workshop]}, September 2020.
\newblock https://doi.org/10.5281/zenodo.4025092.

\bibitem[Lichko et~al.(2023)Lichko, Endrizzi, Juno, Olson, Dorfman, Young, Thakur, Kostadinova, Abler, Li, Schaeffer, Klein, Filwett, Rivera, Guidoni, Timm, Heuer, TenBarge, Matthews, Arzamasskiy, Du, Howes, Comisso, Effenberg, Fries, Squire, Shi, Mallet, Verniero, Ofman, Meyrand, Moreland, Wang, Adhikari, Ledvina, Dong, Gilly, Ghadjari, Light, Sarkar, Liu, Swisdak, Lynch, Maharana, Fu, Wanliss, Kumar, Cassak, and Kumari]{Lichko2023}
Emily Lichko, Douglass Endrizzi, James Juno, Joseph Olson, Seth Dorfman, Rachel Young, Saikat~Chakraborty Thakur, Evdokiya Kostadinova, Mel Abler, Feiyu Li, Derek Schaeffer, Kristopher Klein, Rachael Filwett, Yeimy Rivera, Silvina Guidoni, Arian Timm, Peter Heuer, Jason TenBarge, Lorin Matthews, Lev Arzamasskiy, Tiger Du, Gregory~G. Howes, Luca Comisso, Florian Effenberg, Dan Fries, Jonathan Squire, Peiyun Shi, Alfred Mallet, Jaye Verniero, Leon Ofman, Romain Meyrand, Kimberly Moreland, Liang Wang, Subash Adhikari, Vincent Ledvina, Chuanfei Dong, Chris Gilly, Hossein Ghadjari, Christopher Light, Ranadeep Sarkar, Yi-Hsin Liu, Marc Swisdak, Benjamin~J. Lynch, Anwesha Maharana, Xiangrong Fu, James Wanliss, Pankaj Kumar, Paul Cassak, and Anshu Kumari.
\newblock Enabling {Discoveries} in {Heliospheric} {Science} through {Laboratory} {Plasma} {Experiments}.
\newblock \emph{Bulletin of the AAS}, 55\penalty0 (3), jul 31 2023.
\newblock https://baas.aas.org/pub/2023n3i236.

\bibitem[Limpaecher and MacKenzie(1973)]{limpaecher1973magnetic}
Rudolf Limpaecher and KR~MacKenzie.
\newblock Magnetic multipole containment of large uniform collisionless quiescent plasmas.
\newblock \emph{Review of Scientific Instruments}, 44\penalty0 (6):\penalty0 726--731, 1973.

\bibitem[Lochhaas et~al.(2023)Lochhaas, Tumlinson, Peeples, O’Shea, Werk, Simons, Juno, Kopenhafer, Augustin, Wright, Acharyya, and Smith]{Lochhaas:2023}
Cassandra Lochhaas, Jason Tumlinson, Molly~S. Peeples, Brian~W. O’Shea, Jessica~K. Werk, Raymond~C. Simons, James Juno, Claire Kopenhafer, Ramona Augustin, Anna~C. Wright, Ayan Acharyya, and Britton~D. Smith.
\newblock Figuring out gas \& galaxies in enzo (foggie). vi. the circumgalactic medium of l\* galaxies is supported in an emergent, nonhydrostatic equilibrium.
\newblock \emph{Astrophys.~J.}, 948\penalty0 (1):\penalty0 43, may 2023.
\newblock \doi{10.3847/1538-4357/acbb06}.
\newblock URL \url{https://dx.doi.org/10.3847/1538-4357/acbb06}.

\bibitem[Magee et~al.(2011)Magee, Den~Hartog, Kumar, Almagri, Chapman, Fiksel, Mirnov, Mezonlin, and Titus]{magee2011anisotropic}
RM~Magee, DJ~Den~Hartog, STA Kumar, AF~Almagri, BE~Chapman, G~Fiksel, VV~Mirnov, ED~Mezonlin, and JB~Titus.
\newblock Anisotropic ion heating and tail generation during tearing mode magnetic reconnection in a high-temperature plasma.
\newblock \emph{Physical Review Letters}, 107\penalty0 (6):\penalty0 065005, 2011.

\bibitem[Major et~al.(2022)Major, McKee, Geiger, Den~Hartog, Jaehnig, Seyfert, Smith, Stewart, and Yan]{major2022pedestal}
MR~Major, GR~McKee, B~Geiger, DJ~Den~Hartog, K~Jaehnig, C~Seyfert, DR~Smith, SD~Stewart, and Z~Yan.
\newblock Pedestal fluctuation measurements with charge exchange imaging at the diii-d tokamak.
\newblock \emph{Review of Scientific Instruments}, 93\penalty0 (11), 2022.

\bibitem[Mallet et~al.(2023)Mallet, Dorfman, Abler, Bowen, and Chen]{mallet23}
Alfred Mallet, Seth Dorfman, Mel Abler, Trevor~A. Bowen, and Christopher H.~K. Chen.
\newblock Nonlinear dynamics of small-scale {{Alfv{\'e}n}} waves.
\newblock \emph{Physics of Plasmas}, 30\penalty0 (11):\penalty0 112102, November 2023.
\newblock ISSN 1070-664X.
\newblock \doi{10.1063/5.0151035}.

\bibitem[McKee et~al.(1999)McKee, Ashley, Durst, Fonck, Jakubowski, Tritz, Burrell, Greenfield, and Robinson]{mckee1999beam}
G~McKee, R~Ashley, R~Durst, R~Fonck, Marcin Jakubowski, K~Tritz, K~Burrell, Charles Greenfield, and John Robinson.
\newblock The beam emission spectroscopy diagnostic on the diii-d tokamak.
\newblock \emph{Review of scientific instruments}, 70\penalty0 (1):\penalty0 913--916, 1999.

\bibitem[Meinecke et~al.(2022)Meinecke, Tzeferacos, Ross, Bott, Feister, Park, Bell, Blandford, Berger, Bingham, Casner, Chen, Foster, Froula, Goyon, Kalantar, Koenig, Lahmann, Li, Lu, Palmer, Petrasso, Poole, Remington, Reville, Reyes, Rigby, Ryu, Swadling, Zylstra, Miniati, Sarkar, Schekochihin, Lamb, and Gregori]{Meinecke:2022}
Jena Meinecke, Petros Tzeferacos, James~S. Ross, Archie F.~A. Bott, Scott Feister, Hye-Sook Park, Anthony~R. Bell, Roger Blandford, Richard~L. Berger, Robert Bingham, Alexis Casner, Laura~E. Chen, John Foster, Dustin~H. Froula, Clement Goyon, Daniel Kalantar, Michel Koenig, Brandon Lahmann, Chikang Li, Yingchao Lu, Charlotte A.~J. Palmer, Richard~D. Petrasso, Hannah Poole, Bruce Remington, Brian Reville, Adam Reyes, Alexandra Rigby, Dongsu Ryu, George Swadling, Alex Zylstra, Francesco Miniati, Subir Sarkar, Alexander~A. Schekochihin, Donald~Q. Lamb, and Gianluca Gregori.
\newblock Strong suppression of heat conduction in a laboratory replica of galaxy-cluster turbulent plasmas.
\newblock \emph{Science Advances}, 8\penalty0 (10):\penalty0 eabj6799, 2022.
\newblock \doi{10.1126/sciadv.abj6799}.
\newblock URL \url{https://www.science.org/doi/abs/10.1126/sciadv.abj6799}.

\bibitem[Milchberg and Scime(2020)]{milchberg20}
Howard Milchberg and Earl Scime.
\newblock Workshop on opportunities, challenges, and best practices for basic plasma science user facilities. final report--conference proposal.
\newblock Technical report, US Department of Energy (USDOE), Washington, DC (United States). Office of Science and Technical Information, 4 2020.
\newblock URL \url{https://www.osti.gov/biblio/1615521}.

\bibitem[{National Academies of Sciences, Engineering, and Medicine}(2024)]{heliodecadal24}
{National Academies of Sciences, Engineering, and Medicine}.
\newblock \emph{The Next Decade of Discovery in Solar and Space Physics: Exploring and Safeguarding Humanity's Home in Space}.
\newblock The National Academies Press, Washington, DC, 2024.
\newblock ISBN 978-0-309-73250-5.
\newblock \doi{10.17226/27938}.
\newblock URL \url{https://nap.nationalacademies.org/catalog/27938/the-next-decade-of-discovery-in-solar-and-space-physics}.

\bibitem[Owen et~al.(2023)Owen, Wu, Inoue, Yang, and Mitchell]{Owen2023}
Ellis~R. Owen, Kinwah Wu, Yoshiyuki Inoue, H.~Y.Karen Yang, and Alison~M.W. Mitchell.
\newblock {Cosmic Ray Processes in Galactic Ecosystems}.
\newblock \emph{Galaxies}, 11\penalty0 (4):\penalty0 1--78, 2023.
\newblock ISSN 20754434.
\newblock \doi{10.3390/galaxies11040086}.

\bibitem[Palmer et~al.(2005)Palmer, Gekelman, and Vincena]{palmer05}
Nathan Palmer, Walter Gekelman, and Stephen Vincena.
\newblock {Measurement of ion motion in a shear Alfvén wave}.
\newblock \emph{Physics of Plasmas}, 12\penalty0 (7):\penalty0 072102, 06 2005.
\newblock ISSN 1070-664X.
\newblock \doi{10.1063/1.1930796}.
\newblock URL \url{https://doi.org/10.1063/1.1930796}.

\bibitem[{Peterson} et~al.(2021){Peterson}, {Endrizzi}, {Clark}, {Egedal}, {Flanagan}, {Loureiro}, {Milhone}, {Olson}, {Sovinec}, {Wallace}, and {Forest}]{Peterson2021}
Ethan~E. {Peterson}, Douglass~A. {Endrizzi}, Michael {Clark}, Jan {Egedal}, Kenneth {Flanagan}, Nuno~F. {Loureiro}, Jason {Milhone}, Joseph {Olson}, Carl~R. {Sovinec}, John {Wallace}, and Cary~B. {Forest}.
\newblock {Laminar and turbulent plasmoid ejection in a laboratory Parker Spiral current sheet}.
\newblock \emph{Journal of Plasma Physics}, 87\penalty0 (4):\penalty0 905870410, August 2021.
\newblock \doi{10.1017/S0022377821000775}.

\bibitem[Quataert(2008)]{Quataert2008}
Eliot Quataert.
\newblock {Buoyancy Instabilities in Weakly Magnetized Low‐Collisionality Plasmas}.
\newblock \emph{The Astrophysical Journal}, 673\penalty0 (2):\penalty0 758--762, 2008.
\newblock ISSN 0004-637X.
\newblock \doi{10.1086/525248}.

\bibitem[Quataert et~al.(2002)Quataert, Dorland, and Hammett]{Quataert2002}
Eliot Quataert, William Dorland, and Gregory~W. Hammett.
\newblock {The Magnetorotational Instability in a Collisionless Plasma}.
\newblock \emph{The Astrophysical Journal}, 577\penalty0 (1):\penalty0 524--533, 2002.
\newblock ISSN 0004-637X.
\newblock \doi{10.1086/342174}.

\bibitem[Rice et~al.(1997)Rice, Burrell, and Lao]{rice1997effect}
BW~Rice, KH~Burrell, and LL~Lao.
\newblock Effect of plasma radial electric field on motional stark effect measurements and equilibrium reconstruction.
\newblock \emph{Nuclear fusion}, 37\penalty0 (4):\penalty0 517, 1997.

\bibitem[{Richardson} et~al.(2022){Richardson}, {Burlaga}, {Elliott}, {Kurth}, {Liu}, and {von Steiger}]{Richardson:2022}
J.~D. {Richardson}, L.~F. {Burlaga}, H.~{Elliott}, W.~S. {Kurth}, Y.~D. {Liu}, and R.~{von Steiger}.
\newblock {Observations of the Outer Heliosphere, Heliosheath, and Interstellar Medium}.
\newblock \emph{Space Sci.~Rev.}, 218\penalty0 (4):\penalty0 35, June 2022.
\newblock \doi{10.1007/s11214-022-00899-y}.

\bibitem[Riquelme et~al.(2016)Riquelme, Quataert, and Verscharen]{Riquelme:2016}
Mario~A. Riquelme, Eliot Quataert, and Daniel Verscharen.
\newblock Pic simulations of the effect of velocity space instabilities on electron viscosity and thermal conduction.
\newblock \emph{Astrophys.~J.}, 824\penalty0 (2):\penalty0 123, jun 2016.
\newblock \doi{10.3847/0004-637X/824/2/123}.
\newblock URL \url{https://dx.doi.org/10.3847/0004-637X/824/2/123}.

\bibitem[Roberg-Clark et~al.(2016)Roberg-Clark, Drake, Reynolds, and Swisdak]{RobergClark:2016}
G.~T. Roberg-Clark, J.~F. Drake, C.~S. Reynolds, and M.~Swisdak.
\newblock Suppression of electron thermal conduction in the high $\beta$ intracluster medium of galaxy clusters.
\newblock \emph{Astrophys.~J.~Lett.}, 830\penalty0 (1):\penalty0 L9, oct 2016.
\newblock \doi{10.3847/2041-8205/830/1/L9}.
\newblock URL \url{https://dx.doi.org/10.3847/2041-8205/830/1/L9}.

\bibitem[Roberg-Clark et~al.(2018)Roberg-Clark, Drake, Reynolds, and Swisdak]{RobergClark:2018}
G.~T. Roberg-Clark, J.~F. Drake, C.~S. Reynolds, and M.~Swisdak.
\newblock Suppression of electron thermal conduction by whistler turbulence in a sustained thermal gradient.
\newblock \emph{Phys.~Rev.~Lett.}, 120:\penalty0 035101, Jan 2018.
\newblock \doi{10.1103/PhysRevLett.120.035101}.
\newblock URL \url{https://link.aps.org/doi/10.1103/PhysRevLett.120.035101}.

\bibitem[Rosin et~al.(2011)Rosin, Schekochihin, Rincon, and Cowley]{Rosin:2011}
M.~S. Rosin, A.~A. Schekochihin, F.~Rincon, and S.~C. Cowley.
\newblock {A non-linear theory of the parallel firehose and gyrothermal instabilities in a weakly collisional plasma}.
\newblock \emph{Monthly Notices of the Royal Astronomical Society}, 413\penalty0 (1):\penalty0 7--38, 04 2011.
\newblock ISSN 0035-8711.
\newblock \doi{10.1111/j.1365-2966.2010.17931.x}.
\newblock URL \url{https://doi.org/10.1111/j.1365-2966.2010.17931.x}.

\bibitem[Schaffner et~al.(2014)Schaffner, Wan, and Brown]{Schaffner2014}
D.~A. Schaffner, A.~Wan, and M.~R. Brown.
\newblock Observation of turbulent intermittency scaling with magnetic helicity in an mhd plasma wind tunnel.
\newblock \emph{Phys. Rev. Lett.}, 112:\penalty0 165001, Apr 2014.
\newblock \doi{10.1103/PhysRevLett.112.165001}.
\newblock URL \url{https://link.aps.org/doi/10.1103/PhysRevLett.112.165001}.

\bibitem[Schekochihin et~al.(2008)Schekochihin, Cowley, Kulsrud, Rosin, and Heinemann]{Schekochihin:2008a}
A.~A. Schekochihin, S.~C. Cowley, R.~M. Kulsrud, M.~S. Rosin, and T.~Heinemann.
\newblock Nonlinear growth of firehose and mirror fluctuations in astrophysical plasmas.
\newblock \emph{Phys.~Rev.~Lett.}, 100:\penalty0 081301, Feb 2008.
\newblock \doi{10.1103/PhysRevLett.100.081301}.
\newblock URL \url{https://link.aps.org/doi/10.1103/PhysRevLett.100.081301}.

\bibitem[Schroeder et~al.(2021)Schroeder, Howes, Kletzing, Skiff, Carter, Vincena, and Dorfman]{Schroeder2021}
J.~W.R. Schroeder, G.~G. Howes, C.~A. Kletzing, F.~Skiff, T.~A. Carter, S.~Vincena, and S.~Dorfman.
\newblock Laboratory measurements of the physics of auroral electron acceleration by alfvén waves.
\newblock \emph{Nature Communications}, 12:\penalty0 1--9, 2021.
\newblock ISSN 20411723.
\newblock \doi{10.1038/s41467-021-23377-5}.
\newblock URL \url{http://dx.doi.org/10.1038/s41467-021-23377-5}.

\bibitem[Scime(2024)]{test1}
Earl Scime.
\newblock Private communication, 2024.

\bibitem[Shah et~al.(1999)Shah, Connor, Lei, Schoch, Crowley, Schatz, and Dong]{shah1999heavy}
U~Shah, KA~Connor, J~Lei, PM~Schoch, TP~Crowley, JG~Schatz, and Y~Dong.
\newblock A heavy ion beam probe for the madison symmetric torus.
\newblock \emph{Review of scientific instruments}, 70\penalty0 (1):\penalty0 963--966, 1999.

\bibitem[Shi and Scime(2023)]{shi2023multi}
Peiyun Shi and Earl~E Scime.
\newblock Multi-dimensional incoherent thomson scattering system in phase space mapping (phasma) facility.
\newblock \emph{Review of Scientific Instruments}, 94\penalty0 (2), 2023.

\bibitem[Shi et~al.(2022)Shi, Srivastav, Barbhuiya, Cassak, Scime, and Swisdak]{shi2022evdf}
Peiyun Shi, Prabhakar Srivastav, M.~Hasan Barbhuiya, Paul~A. Cassak, Earl~E. Scime, and M.~Swisdak.
\newblock Laboratory observations of electron heating and non-maxwellian distributions at the kinetic scale during electron-only magnetic reconnection.
\newblock \emph{Phys. Rev. Lett.}, 128:\penalty0 025002, Jan 2022.
\newblock \doi{10.1103/PhysRevLett.128.025002}.
\newblock URL \url{https://link.aps.org/doi/10.1103/PhysRevLett.128.025002}.

\bibitem[Spitkovsky(2008)]{Spitkovsky2008}
Anatoly Spitkovsky.
\newblock {Particle Acceleration in Relativistic Collisionless Shocks: Fermi Process at Last?}
\newblock \emph{The Astrophysical Journal}, 2008.
\newblock ISSN 0004-637X.
\newblock \doi{10.1086/590248}.

\bibitem[Talbot et~al.(2024)Talbot, Pakmor, Pfrommer, Springel, Werhahn, Bieri, and van~de Voort]{talbot2024}
Rosie~Y. Talbot, Rüdiger Pakmor, Christoph Pfrommer, Volker Springel, Maria Werhahn, Rebekka Bieri, and Freeke van~de Voort.
\newblock Anisotropic thermal conduction on a moving mesh for cosmological simulations, 2024.
\newblock URL \url{https://arxiv.org/abs/2410.07316}.

\bibitem[Thakur et~al.(2014{\natexlab{a}})Thakur, Brandt, Cui, Gosselin, Light, and Tynan]{thakur2014multi}
S~Chakraborty Thakur, C~Brandt, L~Cui, JJ~Gosselin, AD~Light, and GR~Tynan.
\newblock Multi-instability plasma dynamics during the route to fully developed turbulence in a helicon plasma.
\newblock \emph{Plasma Sources Science and Technology}, 23\penalty0 (4):\penalty0 044006, 2014{\natexlab{a}}.

\bibitem[Thakur et~al.(2014{\natexlab{b}})Thakur, Brandt, Light, Cui, Gosselin, and Tynan]{thakur2014simultaneous}
SC~Thakur, C~Brandt, A~Light, L~Cui, JJ~Gosselin, and GR~Tynan.
\newblock Simultaneous use of camera and probe diagnostics to unambiguously identify and study the dynamics of multiple underlying instabilities during the route to plasma turbulence.
\newblock \emph{Review of Scientific Instruments}, 85\penalty0 (11), 2014{\natexlab{b}}.

\bibitem[{The Event Horizon Telescope Collaboration} et~al.(2019{\natexlab{a}})]{EHT:2019a}
{The Event Horizon Telescope Collaboration} et~al.
\newblock First m87 event horizon telescope results. i. the shadow of the supermassive black hole.
\newblock \emph{Astrophys.~J.~Lett.}, 875\penalty0 (1):\penalty0 L1, apr 2019{\natexlab{a}}.
\newblock \doi{10.3847/2041-8213/ab0ec7}.
\newblock URL \url{https://dx.doi.org/10.3847/2041-8213/ab0ec7}.

\bibitem[{The Event Horizon Telescope Collaboration} et~al.(2019{\natexlab{b}})]{EHT:2019b}
{The Event Horizon Telescope Collaboration} et~al.
\newblock {First M87 Event Horizon Telescope Results. VI. The Shadow and Mass of the Central Black Hole}.
\newblock \emph{Astrophys.~J.~Lett.}, 875\penalty0 (1):\penalty0 L6, apr 2019{\natexlab{b}}.
\newblock \doi{10.3847/2041-8213/ab1141}.
\newblock URL \url{https://dx.doi.org/10.3847/2041-8213/ab1141}.

\bibitem[{The Event Horizon Telescope Collaboration} et~al.(2021)]{EHT:2021}
{The Event Horizon Telescope Collaboration} et~al.
\newblock {First M87 Event Horizon Telescope Results. VIII. Magnetic Field Structure near The Event Horizon}.
\newblock \emph{Astrophys.~J.~Lett.}, 910\penalty0 (L13):\penalty0 43, 2021.
\newblock URL \url{https://iopscience.iop.org/article/10.3847/2041-8213/abe4de}.

\bibitem[Thomas and Pfrommer(2022)]{Thomas2022}
T~Thomas and C~Pfrommer.
\newblock {Comparing different closure relations for cosmic ray hydrodynamics}.
\newblock \emph{Mon. Not. R. Astron. Soc.}, 4816:\penalty0 4803--4816, 2022.

\bibitem[Thuecks et~al.(2009)Thuecks, Kletzing, Skiff, Bounds, and Vincena]{thuecks2009tests}
DJ~Thuecks, CA~Kletzing, F~Skiff, SR~Bounds, and S~Vincena.
\newblock Tests of collision operators using laboratory measurements of shear alfv{\'e}n wave dispersion and damping.
\newblock \emph{Physics of plasmas}, 16\penalty0 (5), 2009.

\bibitem[{Tu} and {Marsch}(1995)]{Tu:1995}
C.-Y. {Tu} and E.~{Marsch}.
\newblock {MHD structures, waves and turbulence in the solar wind: Observations and theories}.
\newblock \emph{Space Sci.~Rev.}, 73:\penalty0 1--2, July 1995.
\newblock \doi{{10.1007/BF00748891}}.

\bibitem[Tumlinson et~al.(2017)Tumlinson, Peeples, and Werk]{Tumlinson:2017}
Jason Tumlinson, Molly~S. Peeples, and Jessica~K. Werk.
\newblock The circumgalactic medium.
\newblock \emph{Ann.~Rev.~Astron.~Astrophys.}, 55\penalty0 (1):\penalty0 389--432, 2017.
\newblock \doi{10.1146/annurev-astro-091916-055240}.

\bibitem[Tóth et~al.(2012)Tóth, {van der Holst}, Sokolov, {De Zeeuw}, Gombosi, Fang, Manchester, Meng, Najib, Powell, Stout, Glocer, Ma, and Opher]{Toth2012}
Gábor Tóth, Bart {van der Holst}, Igor~V. Sokolov, Darren~L. {De Zeeuw}, Tamas~I. Gombosi, Fang Fang, Ward~B. Manchester, Xing Meng, Dalal Najib, Kenneth~G. Powell, Quentin~F. Stout, Alex Glocer, Ying-Juan Ma, and Merav Opher.
\newblock Adaptive numerical algorithms in space weather modeling.
\newblock \emph{Journal of Computational Physics}, 231\penalty0 (3):\penalty0 870--903, 2012.
\newblock ISSN 0021-9991.
\newblock \doi{https://doi.org/10.1016/j.jcp.2011.02.006}.
\newblock URL \url{https://www.sciencedirect.com/science/article/pii/S002199911100088X}.
\newblock Special Issue: Computational Plasma Physics.

\bibitem[Van~Zeeland and Carlstrom(2004)]{van2004phase}
MA~Van~Zeeland and TN~Carlstrom.
\newblock Phase error correction method for a vibration compensated interferometer.
\newblock \emph{Review of scientific instruments}, 75\penalty0 (10):\penalty0 3423--3425, 2004.

\bibitem[Verscharen et~al.(2019)Verscharen, Klein, and Maruca]{Verscharen:2019}
Daniel Verscharen, Kristopher~G. Klein, and Bennett~A. Maruca.
\newblock The multi-scale nature of the solar wind.
\newblock \emph{Living Rev.~Solar Phys.}, 16\penalty0 (1):\penalty0 5, 2019.
\newblock \doi{10.1007/s41116-019-0021-0}.
\newblock URL \url{https://doi.org/10.1007/s41116-019-0021-0}.

\bibitem[Wilson et~al.(2014)Wilson, Sibeck, Breneman, Contel, Cully, Turner, Angelopoulos, and Malaspina]{Wilson2014}
L.~B. Wilson, D.~G. Sibeck, A.~W. Breneman, O.~Le Contel, C.~Cully, D.~L. Turner, V.~Angelopoulos, and D.~M. Malaspina.
\newblock {Quantified energy dissipation rates in the terrestrial bow shock: 2. Waves and dissipation}.
\newblock \emph{Journal of Geophysical Research: Space Physics}, 119\penalty0 (8):\penalty0 6475--6495, 2014.
\newblock ISSN 2169-9380.
\newblock \doi{10.1002/2014ja019930}.

\bibitem[Wilson et~al.(2016)Wilson, Sibeck, Turner, Osmane, Caprioli, and Angelopoulos]{Wilson2016}
L.~B. Wilson, D.~G. Sibeck, D.~L. Turner, A.~Osmane, D.~Caprioli, and V.~Angelopoulos.
\newblock {Relativistic electrons produced by foreshock disturbances observed upstream of earth's bow shock}.
\newblock \emph{Physical Review Letters}, 2016.
\newblock ISSN 10797114.
\newblock \doi{10.1103/PhysRevLett.117.215101}.

\bibitem[Zhang et~al.(2008)Zhang, Heidbrink, Boehmer, McWilliams, Chen, Breizman, Vincena, Carter, Leneman, Gekelman, et~al.]{zhang2008spectral}
Yang Zhang, WW~Heidbrink, H~Boehmer, R~McWilliams, Guangye Chen, BN~Breizman, S~Vincena, T~Carter, D~Leneman, W~Gekelman, et~al.
\newblock Spectral gap of shear alfv{\'e}n waves in a periodic array of magnetic mirrors.
\newblock \emph{Physics of Plasmas}, 15\penalty0 (1), 2008.

\end{thebibliography}

\end{document}